\documentclass[nonacm,sigplan]{acmart}
\usepackage{amsmath}
\usepackage{algorithmic}
\usepackage{graphicx}
\usepackage{textcomp}
\usepackage{xcolor}
\usepackage{fancyhdr}
\usepackage{float}
\usepackage{multirow}

\usepackage{enumitem}

\usepackage{xspace}

\usepackage{cleveref}

\usepackage[cachedir=minted-cache, frozencache=true]{minted}

\setminted[c]{breaklines,xleftmargin=2em,linenos,tabsize=4,obeytabs,frame=single,framesep=5pt}

\newcommand{\parhead}[1]{\vspace{0pt plus 1pt}\noindent{\textbf{#1.}}}
\ifdefined\submissionready
\newcommand{\colorcomment}[2]{\leavevmode\unskip\relax}
\else
\newcommand{\colorcomment}[2]{\leavevmode\unskip\space{\color{#1}#2}\xspace}
\fi

\definecolor{darkviolet}{HTML}{9400D3}
\definecolor{PineGreen}{HTML}{01796F}
\definecolor{neonpink}{HTML}{FF10F0}

\newcommand{\ignore}[1]{}

\newcommand{\rh}{Rowhammer }
\newcommand{\rhend}{Rowhammer}
\newcommand{\TRH}{$T_{RH}\ $}
\newcommand{\TRHEND}{$T_{RH}$}

\newcommand{\RFMab}{\texttt{RFMab}\ }
\newcommand{\RFMabEND}{\texttt{RFMab}}

\newcommand{\RFMsb}{\texttt{RFMsb}\ }
\newcommand{\RFMsbEND}{\texttt{RFMsb}}

\newcommand{\RAAMMT}{\texttt{RAAMMT}\ }
\newcommand{\RAAMMTEND}{\texttt{RAAMMT}}

\newcommand{\RAAIMT}{\texttt{RAAIMT}\ }
\newcommand{\RAAIMTEND}{\texttt{RAAIMT}}

\newcommand{\RAACtr}{\texttt{RAACtr}\ }
\newcommand{\RAACtrEND}{\texttt{RAACtr}}

\newcommand{\REF}{\texttt{REF}\ }
\newcommand{\REFEND}{\texttt{REF}}

\newcommand{\ACT}{\texttt{ACT}\ }

\newcommand{\tREFI}{\texttt{tREFI}\ }
\newcommand{\tREFIEND}{\texttt{tREFI}}

\newcommand{\tREFIs}{\texttt{tREFIs}\ }
\newcommand{\tREFIsEND}{\texttt{tREFIs}}

\newcommand{\tREFWEND}{\texttt{tREFW}}

\newcommand{\tRFC}{\texttt{tRFC}\ }
\newcommand{\tRFCEND}{\texttt{tRFC}}

\newcommand{\tRC}{\texttt{tRC}\ }
\newcommand{\tRCEND}{\texttt{tRC}}

\begin{document}
\title{RogueRFM: Attacking Refresh Management\\ for Covert-Channel and Denial-of-Service}


\author{Hritvik Taneja}
\affiliation{
  \institution{Georgia Institute of Technology}
  \country{}
}

\author{Moinuddin Qureshi}
\affiliation{
  \institution{Georgia Institute of Technology}
  \country{}
}


\begin{abstract}
With lowering thresholds, transparently defending against Rowhammer within DRAM is challenging due to the lack of time to perform mitigation. Commercially deployed in-DRAM defenses like TRR that steal time from normal refreshes~(REF) to perform mitigation have been proven ineffective against Rowhammer. In response, a new {\em Refresh Management (RFM)} interface has been added to the DDR5 specifications. RFM provides dedicated time to an in-DRAM defense to perform mitigation. Several recent works have used RFM for the intended purpose -- building better Rowhammer defenses. However, to the best of our knowledge, no prior study has looked at the potential security implications of this new feature if an attacker subjects it to intentional misuse.

Our paper shows that RFM introduces new side effects in the system - the activity of one bank causes interference with the operation of the other banks. Thus, the latency of a bank becomes dependent on the activity of other banks. We use these side effects to build two new attacks.  First, a novel {\em memory-based covert channel}, which has a bandwidth of up to 31.3 KB/s, and is also effective even in a bank-partitioned system. Second, a new {\em Denial-of-Service (DOS)} attack pattern that exploits the activity within a single bank to reduce the performance of the other banks. Our experiments on SPEC2017, PARSEC, and LIGRA workloads show a slowdown of up to 67\% when running alongside our DOS pattern. We also discuss potential countermeasures for our attacks. 
\end{abstract}
\maketitle 
\pagestyle{plain} 

\section{Introduction}
\label{sec:introduction}

\rhend~\cite{kim2014flipping} is a phenomenon where repeated activations to a DRAM row can cause bit flips in neighboring rows. An attacker can exploit \rh to flip bits across security boundaries, such as protected page table entries~\cite{seaborn2015exploiting}, leading to a wide range of security vulnerabilities~\cite{gruss2016rhjs,vanderveen2016drammer,gruss2018another,cojocar2019eccploit,frigo2020trrespass,kwong2020rambleed,fahr2022frodo}. \emph{\rh Threshold (\TRHEND)}, the minimum number of activations required to induce a bit flip using \rh has reduced from $139K$~\cite{kim2014flipping} to $4.8K$~\cite{kim2020revisitingRH} over the last decade. \TRH is used to characterize \rhend. This alarming reduction of \TRH has highlighted the need for defenses that would be effective even at low  \TRHEND, which are expected to be present in the near future.


To that end, a number of hardware-based \rh defenses have been proposed~\cite{qureshi2022hydra,saxena2024start,marazzi2022protrr,qureshi2024mint,park2020graphene,kim2022mithril,jaleel2024pride}. A typical hardware-based \rh defense \emph{tracks} frequently activated \emph{attacker rows} and performs \emph{mitigative action} to neighboring \emph{victim rows} if the number of activations to the attacker rows is likely to reach \TRHEND. The typical mitigative action performs a \emph{refresh} operation to the victim rows. Tracking frequently activated rows can be done on the memory controller or within the DRAM itself. The in-DRAM approach is appealing to the DRAM vendors because it can transparently protect against \rh and allows the vendors to tailor their solutions to the properties of their chips.


\parhead{The Need for Time}
With lowering \TRHEND, the time needed by an in-DRAM defense to mitigate victim rows increases. This happens due to the increased number of rows breaching \TRHEND. Commercially deployed in-DRAM trackers like \emph{Targeted Row Refresh} (TRR)~\cite{hassan2021uncovering} transparently mitigate victim rows by stealing some fraction of the time from the normal refresh commands. However, this approach of stealing time from \REF is not scalable and limits the maximum tolerable \TRH of a few thousand~\cite{qureshi2024mint,jaleel2024pride,kim2022mithril}. Unsurprisingly, TRR has shown to be ineffective against recent \rh attacks~\cite{frigo2020trrespass,jattke2022blacksmith}. DDR5 introduced the {\em Refresh Management (RFM)} feature~\cite{JEDEC-DDR5,micron_ddr5}, which provides explicit time to do in-DRAM mitigations.

\parhead{Refresh Management (RFM)}
The RFM interface adds the \RFMab and \RFMsb commands to the DRAM protocol for explicit mitigative action against \rhend. An in-DRAM defense refreshes victim rows when any one of the RFM commands is issued by the memory controller. The RFM interface also introduces a mechanism to force the memory controller to launch these commands regularly. Recent works~\cite{marazzi2022protrr,kim2022mithril,jaleel2024pride,qureshi2024mint} have used the RFM interface to build in-DRAM defenses that can tolerate low thresholds.

\parhead{In-DRAM Mitigations Using RFM}
ProTRR and Mithril~\cite{marazzi2022protrr,kim2022mithril} are two provably secure, counter-based in-DRAM trackers that use the RFM interface to perform mitigative actions. Both these schemes build a per-bank tracker that stores an optimal number of counters to track activations for a given \TRHEND. Next, they use these counter values to perform mitigative refreshes to the victim rows via RFM. The prohibitively large SRAM budget for these optimal trackers has inspired the development of low-cost probabilistic trackers. PrIDE~\cite{jaleel2024pride} and MINT~\cite{qureshi2024mint} are two recent works that build a probabilistic tracker to identify the aggressor rows and use the RFM interface to perform mitigative actions. 


\begin{figure*}[htb]
    \centering
    \includegraphics[width=0.999\textwidth]{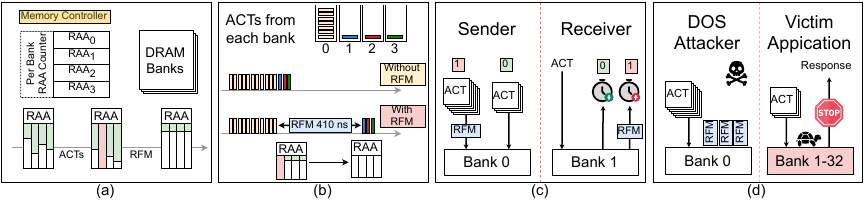}
    \vspace{-2em}
    \caption{(a) Overview of RFM: \RAACtr is incremented on activations, and when \RAACtrEND=\RAAMMTEND, the MC issues an \RFMab command. (b) As \RFMab blocks all banks for 410 ns, allowing one bank to impact the performance of others. (c) A sender can trigger \RFMab to slow all banks, enabling a timing-based covert channel. (d) Continuous ACTs to one bank can cause frequent \RFMab operations, leading to a DOS-like scenario.}
    \Description{Overview}
    \vspace{-1em}
    \label{fig:overview}
\end{figure*}


\parhead{Goal of the Paper}
Prior works have used RFM in the intended manner, to build effective Rowhammer mitigation solutions. However, to the best of our knowledge, RFM has been used only for benign purposes, ignoring any potential misuse. To that end, the goal of our work is to analyze the implications of an RFM-enabled DRAM on system security. In this work, we show that RFM introduces new side effects that can have serious security implications. To understand these negative side effects of RFM, we first explain how a memory controller issues an RFM command.

\parhead{Mechanism to Launch RFM commands}
The RFM interface specifies the memory controller to store a per bank \emph{Rolling Accumulated ACT counter} (\RAACtrEND), which is incremented at every activation. Next, once the \RAACtr reaches a \emph{Maximum Manageable Threshold} (\RAAMMTEND), the memory controller must issue an RFM command. This upper bound on the value of the \RAACtr ensures that the in-DRAM defense has the opportunity to perform mitigative actions at regular intervals. When an RFM command is issued, the memory controller also reduces the value of the \RAACtrEND, allowing further activations to the bank~(\Cref{fig:overview}(a)). The \RFMab command performs mitigation across all the banks in a rank and reduces the \RAACtr of all the banks by a vendor-specified value. In this work, we focus on the \RFMab command.

\parhead{Key Insight}
During an \RFMab command, all the banks in a rank are blocked for 410ns to perform a mitigative action. Next, an \RFMab command is issued even if the \RAACtr of only one bank has reached \RAAMMTEND. This means that a single bank is allowed to block all the banks in a rank for 410ns~(\Cref{fig:overview}(b)). Conversely, the DRAM timings of a bank are now also dependent on the activity of other banks. Thus, RFM enables a bank to interfere with the operation of the other. We show an attacker can misuse this ability to induce slowdowns in co-running applications and to build a covert-channel.

\parhead{RFM-Based Covert Channel}
First, we exploit the RFM interface to develop a new memory-based covert channel between two banks~(\Cref{fig:overview}(c)). The receiver regularly issues loads that trigger ACTs in one of the banks and monitors the time to perform the loads. To send a 1, the sender invokes an access pattern that triggers a large number of ACTs to a different bank, causing its \RAACtr to saturate. The memory controller then issues an \RFMab command, causing a delay in the time taken to perform the loads across all the banks, which the receiver can detect. To send a 0, the sender does trigger any ACTs, resulting in a shorter response time for the receiver, which it can infer as a 0. This covert channel achieves a bandwidth of 31.3 KB/s per sub-channel, resulting in a total bandwidth of 62.6 KB/s per channel. We observe an accuracy of 100\% in the noiseless setting. When evaluated with 1 and 2 additional workloads to add noise, the covert channel maintains an accuracy of 85\% and 83\%, respectively.

\parhead{DOS Attack using RFM}
Next, we use RFM to build a DOS attack~(\Cref{fig:overview}(d)). The attacker traverses an access pattern that issues a large number of ACTs to one of the banks, causing the \RAACtr to reach the \RAAMMTEND. As a result, the memory controller issues an \RFMab command, which blocks all the banks for around 410ns. We find that a DOS attacker that continuously triggers ACTs to one bank can trigger 1.37 \RFMabEND/\tREFI for \RAAMMTEND=96 and 2.6 \RFMabEND/\tREFI for \RAAMMTEND=48. Our evaluation when running the DOS attack pattern alongside SPEC2017, PARSEC, and LIGRA workloads shows a maximum slowdown of 29.3 and 67\% for \RAAMMTEND=96 and 48, respectively. On average, we see a slowdown of 14.7\% and 27.7\% with \RAAMMTEND=96 and 48, respectively.

\parhead{Countermeasures}
Memory isolation techniques like bank and bus partitioning~\cite{saltaformaggio2013busmonitor}, used to mitigate existing memory-based covert channels, render ineffective against our covert channel and the DOS attack. We discuss potential countermeasures to our attacks. We evaluate activation limiting as a countermeasure against the DOS attack. Our evaluation shows that activation limiting can reduce the average slowdown caused by the DOS attack to 10.2\% and 14.2\% with \RAAMMTEND=96 and 48, respectively. Overall, our paper makes the following contributions:
\begin{enumerate}
    \item To the best of our knowledge, this is the first work to look at the security implications of RFM arising from the interference of one bank on the timing of another.  
    \item We use RFM to build a new {\em memory-based covert channel} that is both fast and robust against existing memory isolation techniques such as bank partitioning. 
    \item We use RFM to build a {\em Denial of Service (DOS)} attack that can cause significant slowdowns on co-running applications, even if they are in different banks.

    \item We discuss the potential countermeasures against our covert channel and DOS attack.
\end{enumerate}


\clearpage

\section{Background and Motivation}

\label{sec:background}

\subsection{Threat Model}
\label{sec:threat_model}
\parhead{Covert Channel}
We assume that the sender and receiver processes are running on the same system. The sender wants to transmit data to the receiver using a covert channel. Both the sender and the receiver are aware of the cache indexing function and the DRAM address mapping function. Next, they can evict any cache line using eviction sets~\cite{qureshi2019new,vila2019theory} or \texttt{clflush}~\cite{IntelSDM}. Finally, the sender and receiver are aware of the RFM scheduling policy employed by the memory controller.

\parhead{Denial-of-Service}
The DOS attacker is running on a shared system and is trying to slow down coresident processes or crash the system. The DOS attacker also has the ability to evict any cache line from the caches.

\subsection{DRAM Architecture}
\label{sec:dram_arch}
\parhead{Organization and Timing}
DRAM has a hierarchical structure organized into channels, sub-channels, ranks, banks, and rows. A typical DDR5 configuration has a 64-bit channel containing two subchannels, each 32-bit wide, that operate independently. DDR5 has a burst length of 16 to transfer a 64B cache line. A DDR5 subchannel typically has 16 or 32 banks, each of which stores data in a 2D array of rows and columns. To access data from a bank, an entire row needs to be brought into the \emph{row buffer} using an activation command~(ACT). If the row buffer contains a conflicting row, a \emph{precharge} command closes the row before issuing the \ACT command. The time between two accesses to different rows within the same bank is called the \emph{row cycle time}~(\tRCEND), which is typically around 48ns for DDR5.



\parhead{\REF Command}
DRAM cells store bits as a charge, which they cannot hold indefinitely. So, to ensure data integrity, the memory controller must issue a refresh~(\REFEND) command every \emph{Refresh Interval}~(\tREFIEND=3900ns) to replenish the charge. Each \REF command refreshes a group of rows across multiple banks and takes \tRFCEND~(410ns) time to finish. All the rows are divided into a total of 8192 groups and are refreshed within the refresh window \tREFWEND~(32ms).


\subsection{\rh} 
\rh is a phenomenon that occurs when frequent activations to an aggressor row cause bit flips in the neighboring victim rows. \rh can have severe security implications. \rh has been shown to break confidentiality~\cite{kwong2020rambleed} and perform privilege escalation attacks~\cite{seaborn2015exploiting}.

The minimum number of activations required to induce a bit flip using \rh is called the \emph{\rh Threshold \TRHEND}. Lower \TRH values indicate higher vulnerability to \rhend-based attacks. \TRH has reduced from $139K$~\cite{kim2014flipping} to $4.8K$~\cite{kim2020revisitingRH} over the last decade. This alarming reduction of \TRHEND, fueled by the increasing density of DRAM cells, has highlighted the need for \rh defenses that would work for low thresholds. To that end, numerous \rh defenses have been proposed~\cite{qureshi2022hydra,saxena2024start,marazzi2022protrr,qureshi2024mint,park2020graphene,kim2022mithril,jaleel2024pride}.

\parhead{\rh Defenses}
Typical \rh defenses can be divided into two steps: tracking and mitigation. The tracking step monitors the activations to the rows in a bank and identifies frequently activated \emph{attacker rows}. The mitigation step refreshes the neighbors of the attacker rows called the \emph{victim rows}. This step replenishes the leaked charge from the victim rows, thus preventing bit flips. Tracking the frequently activated rows can be done on the memory controller or within the DRAM itself. The \emph{in-DRAM} approach is appealing as it has two main benefits. First, it can solve \rh within the DRAM without relying on the memory controller while protecting proprietary information about the DRAM implementation. Second, DRAM vendors can tune the \rh defense for each chip based on its \TRHEND. However, the in-DRAM approach suffers from a lack of time to perform the mitigative action.

\parhead{Mitigation Challanges}
Commercially deployed in-DRAM trackers like TRR~\cite{hassan2021uncovering} borrow time from the normal refresh commands (\REFEND) to transparently mitigate victim rows. However, with lowering \TRH values, the number of victim rows increases, and this strategy of stealing time from the \REF command does not scale well. The limited amount of time that can be borrowed from \REF also limits the tolerable \TRH of an in-DRAM defense. Unsurprisingly, it has been shown that TRR is ineffective against recent \rh attacks~\cite{frigo2020trrespass,jattke2022blacksmith}. To address this lack of time for an in-DRAM defense to perform mitigative action, DDR5 introduced the \emph{Refresh Management (RFM)} interface~\cite{JEDEC-DDR5}, which adds a mechanism that provides dedicated time to perform the mitigative action.

\subsection{Refresh Management (RFM)}
\label{sec:rfm}
The RFM interface adds the \RFMab and \RFMsb commands, which, when issued, block the memory controller from issuing further activations until the command finishes execution. During this time, the in-DRAM defense selects an attacker row and refreshes its neighboring victim rows.

\begin{figure}[htb!]
    \centering
    \includegraphics[width=0.85\linewidth]{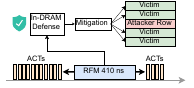}
    \caption{When an RFM command is launched, the memory controller is blocked from issuing activation for 410ns. During this period, the in-DRAM refreshes the victims of an attacker row that has breached \TRHEND.}
    \label{fig:rfm_timeline}
    \Description{RFM Timeline}
\end{figure}

\parhead{RFM Mechanism}
To ensure regular mitigation by the in-DRAM defense, the RFM interface implements a protocol that requires the memory controller to issue an RFM command at fixed intervals. As part of this protocol, the memory controller stores a per bank activation counter called the \emph{Rolling Accumulated ACT Counter} (\RAACtrEND). When the \RAACtr reaches a predefined threshold~(\RAAMMTEND), the memory controller is forced to issue an RFM command. It is important to note that the memory controller can also issue an RFM command before the \RAACtr reaches \RAAMMTEND.

\begin{figure}[htb!]
    \centering
    \includegraphics[width=\linewidth]{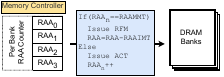}
    \caption{RFM Mechanism: Per-bank \RAACtr is incremented on an activation. If the \RAACtr of any bank reaches the threshold, the MC issues an RFM and reduces the \RAACtrEND.}
    \label{fig:rfm_mechanism}
    \Description{RFM Mechaism}
\end{figure}

\parhead{RFM Commands}
To allow further activations, the RFM commands decrement the \RAACtr by \RAAIMTEND~(specified by DRAM vendor MR58:OP[4:1] bits). The \RFMab (RFM all-bank) command performs mitigation across all the banks and hence reduces the \RAACtr of all the banks within a rank. Next, the \RFMsb (RFM same-bank) command performs mitigation to a specific bank-set~(same bank across all bankgroups) and hence reduces the \RAACtr of a bank-set. In this work, we focus on the \RFMab command. The \RFMab command blocks all the banks within a rank for \tRFC (410 ns).

\parhead{Decrementing \RAACtr in the \REF window}
In addition to the RFM period, the in-DRAM defense can also perform mitigation during REF. Hence, the \RAACtr is also reduced by a fixed amount during REF~(specified by DRAM vendor MR59:OP[7:6] bits). For this work, we assume that during REF, the \RAACtr is decremented by \RAAIMTEND/2.

\parhead{RFM Thresholds}
Recent works~\cite{qureshi2024mint,jaleel2024pride,kim2022mithril} have used \RAAIMT 16 and 32 to tolerate low \TRHEND. So, in this work, we focus on \RAAIMTEND=16 and 32 and \RAAMMTEND=3$\times$\RAAIMTEND. 

\subsection{Effect of RFM on System Security}
\label{sec:rfm_security}
The RFM interface is designed to protect against \rh attack patterns, which typically perform continuous activation to only a few banks. This means that the RFM protocol will launch an RFM command even if the activation counter~(\RAACtrEND) of only 1 of the 32 banks reaches the RFM threshold. As a result, the \RFMab command stalls all the banks for 410ns regardless of which bank triggered RFM.

\begin{figure}[htb!]
    \centering
    \includegraphics[width=\linewidth]{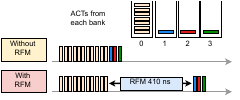}
    \caption{In the presence of RFM, activations from only one bank can interfere with the operation of all the other banks.}
    \label{fig:rfm_adversary}
    \Description{RFM Adversary}
\end{figure}

\parhead{Bank Interference using RFM}
The presence of RFM allows a single bank to affect the operations of all the other banks in the rank. Specifically, continuous activations to one bank will saturate its \RAACtrEND, forcing the memory controller to issue an \RFMabEND, stalling all the banks for 410ns. As a result, any new or queued requests from all the other banks will now have to wait until the RFM command finishes execution. 

\parhead{Side Effects of RFM}
The ability of banks to interfere with the operation of each other introduces two new side effects. \textbf{First}, the DRAM latency of a bank is now affected by the activity of all the other banks in the same rank. This inherently creates a new timing channel, where the response time of a bank is influenced by activity in other banks. \textbf{Second}, RFM allows a single bank to cause slowdowns in the other banks. This opens the door for adversaries to misuse RFM for performance-based attacks. To that end, we now outline the goal of our work.

\begin{figure}[htb!]
    \centering
    \includegraphics[width=\linewidth]{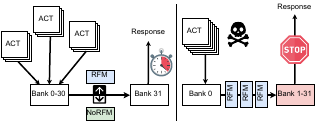}
    \caption{Two side effects of RFM. The latency of one bank is affected by the activity of other banks. One bank has the ability to slow down all the other banks.}
    \label{fig:side_effects}
    \Description{Side Effects of RFM}
\end{figure}

\subsection{Goal: Exploiting Bank Interference from RFM}
RFM is a powerful interface that an in-DRAM defense can use to request time from the memory controller to perform mitigation at regular intervals. Prior works have used RFM for the intended purposes -- building \rh defenses that can tolerate low \TRHEND~\cite{jaleel2024pride,kim2022mithril,marazzi2022protrr}. As with any new feature, it is important to understand the security implications of intentional misuse of the feature.  The goal of our paper is to understand how the interference caused by RFM (whereby the activity of one bank can affect the performance of another) can be used to form new forms of attacks.  


\clearpage

\section{Evaluation Methodology}
\label{sec:methodology}
\subsection{Simulation Framework} 
We use ChampSim~\cite{gober2022championship}, a cycle-level, multi-core, trace-based simulator interfaced with DRAMSim3~\cite{li:dramsim3}, a detailed memory system simulator. We modified DRAMSim3 to include the DDR5 configuration, wherein each DIMM supports two sub-channels that can be operated independently and provides a 64-byte line with a burst length of 16. Table~\ref{table:system_config} shows the configuration for our baseline system.  

\begin {table}[h!]
\begin{footnotesize}
\begin{center} 
\vspace{-0.05 in}
\caption{Baseline System Configuration}
\vspace{-0.05 in}

\begin{tabular}{|c|c|}
\hline
  Out-of-Order Cores             & 4 cores at 4GHz       \\
  ROB size                       & 352       \\
  Fetch, Dispatch, Retire width  & 6, 6, 5         \\ 
  L1-I and L1-D (Private)        & 32KB 8-way and 48KB, 12-way \\
  L2 (Private)                   & 512KB, 8-way \\ \hline
  Last Level Cache (Shared)      & 4MB, 16-Way, 64B lines, LRU \\ \hline
  Memory size                    & 32GB -- DDR5 \\
  Memory bus speed               & 2.4 GHz (4800 MT/s) \\
  Channels                       & 1 (one 32GB DIMM) \\  
  Banks x Ranks x Sub-Channels   & 32$\times$1$\times$2 \\
  Page Size                      & 4KB \\ 
  \hline
\end{tabular}
\label{table:system_config}
\vspace{-0.1 in}
\end{center}
\end{footnotesize}
\end{table}

\noindent We evaluate performance using 4 out-of-order cores with private L1 and L2 caches and shared L3 cache. The L3 is non-inclusive, with 128 MSHRs/core, 128 entry read and write queues, 4 read and write ports, 20-cycle hit-latency, no prefetcher, and an LRU replacement policy. Our memory system contains one channel, with a 32GB DDR5 DIMM. Our memory provisioning of 8GB per core is in line with typical desktops and servers. Next, we assume the system is memory-isolated, as this provides a more secure baseline.

\parhead{Memory Isolation using Bank Partitioning}
We use bank partitioning to build a memory-isolated system, where each core is only allowed access to a particular set of banks. The operating system (OS) allocates pages from only a specific set of banks to each application.  We assume that the OS is aware of the DRAM address mapping function.


\parhead{RFM Policy}
We assume an RFM policy where the memory controller issues an \RFMab command when the \RAACtr of any bank reaches \RAAMMTEND. This reduces the \RAACtr of all the banks within the rank by \RAAIMTEND. We consider two values for \RAAIMTEND, 32 and 16, which implies that \RAAMMT is set to 96 and 48, respectively.

\subsection{Workload Characterization}
\label{sec:wc_characterization}
We evaluate our design using the publicly available ChampSim traces, which include 20 from SPEC2017~\cite{SPEC2017}, 13 from LIGRA~\cite{shun:ligra} (graph processing), and 5 from PARSEC~\cite{bienia2008parsec}. These traces have been collected after fast-forwarding the workload to a region of interest. We perform a warm-up period of 20 million instructions for each workload. Four copies of the same workload run on 4 cores and continue executing until all 4 cores complete 100 million instructions each. 

Table~\ref{table:wc} shows workload characteristics, including the average per-core IPC and LLC-Misses Per 1000 Instructions (MPKI) and the slowdown when running the workload with \RAAIMTEND=32 and \RAAIMTEND=16. We also report the average ACTs per \tREFI across all the 64 banks. The last row shows the average values for \RAAIMTEND=32 and \RAAIMTEND=16.


\begin{table}[!htb]
  \centering
  \begin{footnotesize}
  \caption{Workload Characteristics: IPC, MPKI, ACTs-PKI, and Slowdown with \RAAIMTEND=32 and \RAAIMTEND=16.}
  \label{table:wc}
  \begin{tabular}{|c|c|c|c|c|c|}
    \hline
Workload    & IPC	        &	MPKI	& ACT/tREFI &	\multicolumn{2}{c|}{Slowdown(\%)}	 \\ \cline{5-6}
            & (per-core)	&	(LLC)	& (4-core) &  RFM-32     & RFM-16 \\ \hline \hline
lbm & 0.31 & 36.07 & 7.23 & 0.47 & 9.29 \\
mcf & 0.84 & 26.09 & 14.94 & 3.27 & 17.49 \\
wrf & 0.65 & 20.94 & 3.82 & 0.68 & 8.13 \\
fotonik3d & 0.50 & 19.31 & 10.29 & 1.99 & 15.62 \\
gcc & 1.26 & 17.82 & 0.74 & 0 & 0 \\
omnetpp & 0.62 & 16.97 & 9.71 & 0 & 5.37 \\
bwaves & 0.75 & 13.81 & 10.21 & 9.11 & 29.72 \\
cam4 & 1.25 & 9.13 & 2.30 & 0.01 & 1.57 \\
cactuBSSN & 1.89 & 6.45 & 13.31 & 2.23 & 11.22 \\
roms & 1.72 & 6.13 & 3.66 & 0.63 & 8.16 \\
pop2 & 1.87 & 4.56 & 2.46 & 0.18 & 2.76 \\
xz & 1.80 & 2.84 & 7.76 & 0 & 3.66 \\
xalancbmk & 2.23 & 1.92 & 2.93 & 0 & 0.53 \\
deepsjeng & 2.39 & 0.53 & 1.39 & 0 & 0 \\
x264 & 3.85 & 0.21 & 0.29 & 0 & 0 \\
nab & 3.95 & 0.16 & 0.12 & 0 & 0 \\
perlbench & 3.94 & 0.06 & 0.09 & 0 & 0 \\
imagick & 3.91 & 0.04 & 0.15 & 0 & 0 \\
leela & 3.95 & 0.02 & 0.04 & 0 & 0 \\
exchange2 & 4.00 & 0 & 0 & 0 & 0 \\ \hline \hline
PageRank & 0.35 & 81.44 & 24.14 & 9.15 & 31.75 \\
BellmanFord & 0.61 & 31.63 & 13.45 & 1.11 & 13.07 \\
BC & 0.46 & 26.25 & 1.77 & 0 & 0.74 \\
PR-Delta & 1.38 & 5.99 & 1.72 & 0.02 & 3.09 \\
Triangle & 1.52 & 2.40 & 1.26 & 0 & 2.05 \\
Comp-SC & 1.52 & 2.39 & 1.26 & 0.09 & 2.19 \\
BFSCC & 1.52 & 2.39 & 1.25 & 0 & 2.60 \\
MIS & 1.52 & 2.39 & 1.25 & 0 & 2.60 \\
Radii & 1.52 & 2.39 & 1.25 & 0 & 2.60 \\
Components & 1.52 & 2.38 & 1.25 & 0.07 & 2.23 \\
BFS & 1.52 & 2.38 & 1.26 & 0 & 2.22 \\
BFS-BV & 1.53 & 2.37 & 1.25 & 0.10 & 2.65 \\
CF & 1.72 & 2.01 & 1.19 & 0 & 2.71 \\ \hline \hline
streamc & 1.44 & 13.82 & 2.63 & 0 & 0 \\
facesim & 1.27 & 12.26 & 2.38 & 0.18 & 3.22 \\
canneal & 1.51 & 6.51 & 7.50 & 0 & 4.70 \\
raytrace & 1.17 & 3.45 & 1.79 & 0 & 4.41 \\
fluida & 3.64 & 0.30 & 0.45 & 0 & 0 \\ \hline \hline
Average & 1.13 & 10.15 & 4.17 & 0.68 & 4.93 \\ \hline

  \end{tabular}
 \vspace{-0.1 in}
  \end{footnotesize}
\end{table}

\subsection{Figure of Merit}
For the covert channel, our primary metrics are the accuracy and speed of the channel, measured under conditions where other workloads are running to emulate a noisy system. For the DOS attack, we show the normalized slowdown in the weighted speedup~\cite{eyerman2013restating} of the workloads, when running alongside an adversary with RFM enabled.
\clearpage
\section{Covert Channel using RFM}
\label{sec:covch}
In this section, we use the side effects of RFM to build a memory-based covert channel between two processes in a memory-isolated system~(using bank partitioning). The key insight of our work is that in the presence of the RFM interface, the DRAM timing of a bank can be influenced by the activity of other banks. To transmit `1', a sender process will trigger \RFMab causing a delay in the response time of requests from the receiver process. However, to transmit `0', the sender will not trigger \RFMab, resulting in a shorter response time for the receiver requests. The receiver can detect the difference in the response time of its requests and hence infer the correct data. In the following sections, we describe the access pattern to invoke ACTs, the design of the sender and the receiver, the results of our covert channel, and finally discuss the potential countermeasures.

\begin{figure}[htb!]
    \centering
    \includegraphics[width=\linewidth]{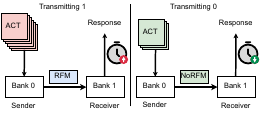}
    \caption{Covert Channel: The RFM interface allows two parties to communicate via a timing channel.}
    \label{fig:covch}
    \Description{Covert Channel}
\end{figure}

\subsection{Access Pattern to Invoke ACTs}
\begin{figure}[htb]
    \centering
    \includegraphics[width=\linewidth]{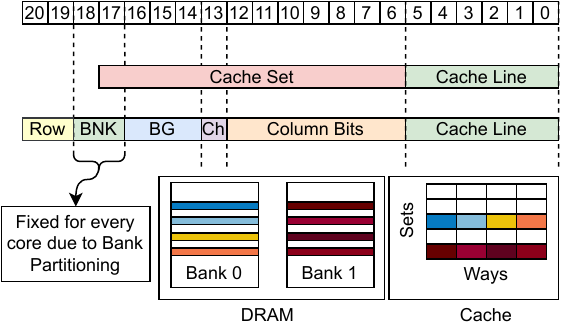}
    \caption{The bits that determine the cache set index also determine the bank index in DRAM. This ensures that all the elements of the eviction set reside in the same bank, and every access to it results in an ACT in the same bank.}
    \label{fig:setup}
    \Description{Self Eviction}
\end{figure}

We want both the sender and the receiver to be able to invoke ACTs to a particular bank. To do that, we \textbf{first} ensure that the data is not served from any cache. To achieve this, we generate a group of addresses that map to the same set in the LLC (eviction set)~\cite{vila2019theory,qureshi2019new}. We create an eviction set of a size greater than the associativity of the LLC. Doing this ensures that every load to the eviction set in a fixed order results in an LLC miss. This oversized eviction set is called a self-evicting set. \textbf{Second}, we rely on the DRAM and the LLC address mapping functions to ensure that all the LLC misses from accessing the eviction set go to the same bank in DRAM. Specifically, we use the fact that the physical address bits that determine the cache set index also determine the bank index in DRAM~(see \Cref{fig:setup}). As a result, every element of the self-evicting set will map to a different row in the same bank. So, every load to the eviction set will result in an ACT.  

\subsection{Design: Sender and Receiver Gadgets}
In this section, we describe the sender and receiver gadgets and show how they can be used to transmit data via a timing channel. For brevity, we assume \RAAIMTEND=32 and \RAAMMTEND=96 throughout this section.

\parhead{Receiver Gadget}
The goal of the receiver is to continuously perform accesses that result in ACTs to its bank and measure the time taken to complete them. To detect timing differences between these ACTs, the receiver should maximize the number of ACTs it can invoke in order to reduce the effect of noise. However, it should not trigger \RFMabEND. Therefore, we limit the number of ACTs from the receiver to 16/\tREFIEND~(\RAAIMTEND/2), which corresponds to the value by which \RAACtr is decremented at every REF.  The receiver performs 16 loads to the self-evicting set within \tREFI and measures the time taken to complete all the loads. If the total time exceeds a certain threshold, the receiver infers a `1'; otherwise, it infers a `0'.

\begin{listing}[htb!]
\begin{minted}{c}
// ACTs per refresh interval (RAAIMT = 32)
uint32_t recv_acts = RAAIMT / 2;
// Wait cycles (CPU) between 2 ACTs
uint32_t wait_cycles = 825;

// Eviction set (to Bank 0)
uint64_t evset[32];
uint32_t counter = 0;

start = rdtsc();
for (int i = 0; i < recv_acts; i++) {
    load(&evset[(counter++) % len]);
    sleep(wait_cycles);
}
end = rdtsc();

// Threshold for RFM vs No-RFM
if (end - start > THRESHOLD) {
    bit = 1;
} else {
    bit = 0;
}
\end{minted}
    \caption{Receiver Gadget}
    \label{lst:receiver}
\end{listing}

\noindent Finally, to ensure that the receiver does not complete the loads before the sender gets a chance to send a bit, these loads are evenly spread over the \tREFI interval. The receiver gadget is shown in \Cref{lst:receiver}.

\begin{figure}[htb!]
    \centering
    \includegraphics[width=\linewidth]{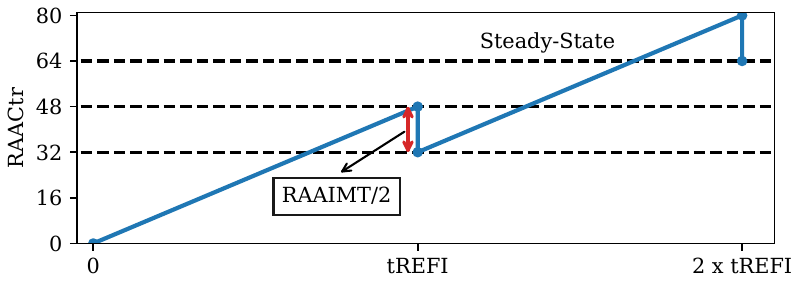}
    \caption{Sender Initialization: The sender issues 96 ACTs to reach the steady state~(\RAACtrEND=64) in 2 \tREFIEND s. After this, the sender can trigger an \RFMab every \tREFIEND.}
    \label{fig:sender_init}
    \Description{Sender Initialization}
\end{figure}

\parhead{Sender Initialization}
The goal of the sender is to transmit a bit every \tREFI by either triggering or not triggering \RFMabEND. To trigger an \RFMab, the \RAACtr must be 96. However, the maximum number of ACTs that can be issued within a \tREFI is approximately 72. Therefore, the sender must undergo an initialization phase, after which it reaches a steady state where it can trigger \RFMab every \tREFIEND. To reach this steady state (\RAACtrEND = 64), the sender issues 48 ACTs/\tREFI to its bank for two consecutive \tREFIsEND. Once in this steady state, the sender can trigger an \RFMab command in subsequent \tREFIs by issuing just 32 ACTs to its bank. \Cref{fig:sender_init} shows the \RAACtr during sender initialization. 

\parhead{Sender Gadget}
To send a `1', the sender issues 48 ACTs to its bank within \tREFIEND. Since the \RAACtr of the sender bank is 64 at the start of \tREFIEND~(steady state), the sender bank triggers an \RFMab command after 32 ACTs. This blocks all the banks in the rank, including the receiver bank, for the next 410 ns~(\tRFCEND). After the banks unblock, the sender issues another 16 ACTs to maintain the steady state. To send a `0', the sender only issues 16 ACTs to its bank within \tREFIEND, which does not trigger RFM but is necessary to maintain the steady state. The sender gadget is shown in \Cref{lst:sender}.

\begin{listing}[htb]
\begin{minted}{c}
// ACTs per refresh interval (RAAIMT = 32)
uint32_t send0_acts = RAAIMT/2; // To send 0
uint32_t send1_acts = RAAIMT + RAAIMT/2; // To send 1

// Wait cycles (CPU) between 2 ACTs
uint32_t wait_cycles = 100; 

// Eviction set (to Bank 1)
uint64_t evset[32];
uint32_t counter = 0;

// Configure number of ACTs and wait cycles
if (send_bit == 0) {
    acts = send0_acts;
} else {
    acts = send1_acts;
}

// Send the bit
for (int i = 0; i < acts; i++) {
    load(&evset[(counter++) % len]);
    sleep(wait_cycles);
}
\end{minted}
    \caption{Sender Gadget}
    \label{lst:sender}
\end{listing}

\begin{figure}[htb]
    \centering
    \includegraphics[width=\linewidth]{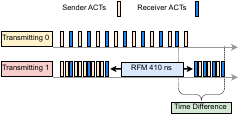}
    \caption{ACT trace when transmitting 0 vs. 1: Frequent ACTs from the sender when transmitting 1 trigger \RFMabEND, resulting in a longer response time for receiver requests.}
    \Description{Timing Difference}
    \label{fig:covch_design}
\end{figure}

\parhead{Timing Difference} 
\Cref{fig:covch_design} shows how the ACTs from the sender and receiver are serviced when transmitting 0 and 1. When sending `1', the sender rapidly issues 48 ACT commands to its bank, triggering an \RFMab command before the receiver finishes its last load. This causes a delay for the receiver. When sending `0', the sender only sends 16 ACTs which does not trigger an \RFMab command, resulting in a shorter response time for the receiver.

\subsection{Refresh Synchronization}
\begin{figure}[htb]
    \centering
    \includegraphics[width=\linewidth]{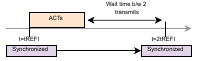}
    \caption{Once synchronized to \tREFIEND, the sender and receiver can be kept aligned to \tREFI by waiting for an appropriate amount of time before the next transmit.}
    \Description{Timing Difference}
    \label{fig:refsync}
\end{figure}

For this covert channel to work, the sender and receiver must be synchronized to the start of the \tREFI interval. This is typically done by continuously accessing two rows in a bank and monitoring the delay in response times between them~\cite{de2021smash,jattke2022blacksmith}. A high delay indicates that a \REF command separates the two requests. Our sender and receiver processes can use this technique to synchronize to the start of the \tREFI interval. Once synchronized, the sender and the receiver keep themselves aligned to \tREFI by waiting for the appropriate time before issuing the requests to transmit the next bit~(see \Cref{fig:refsync}). In our setup, the simulator synchronizes the sender and receiver to \tREFI once at the start of the simulation, and for the rest of the simulation, the sender and receiver synchronize themselves.


\begin{figure*}[htb]
    \centering
    \includegraphics[width=\textwidth]{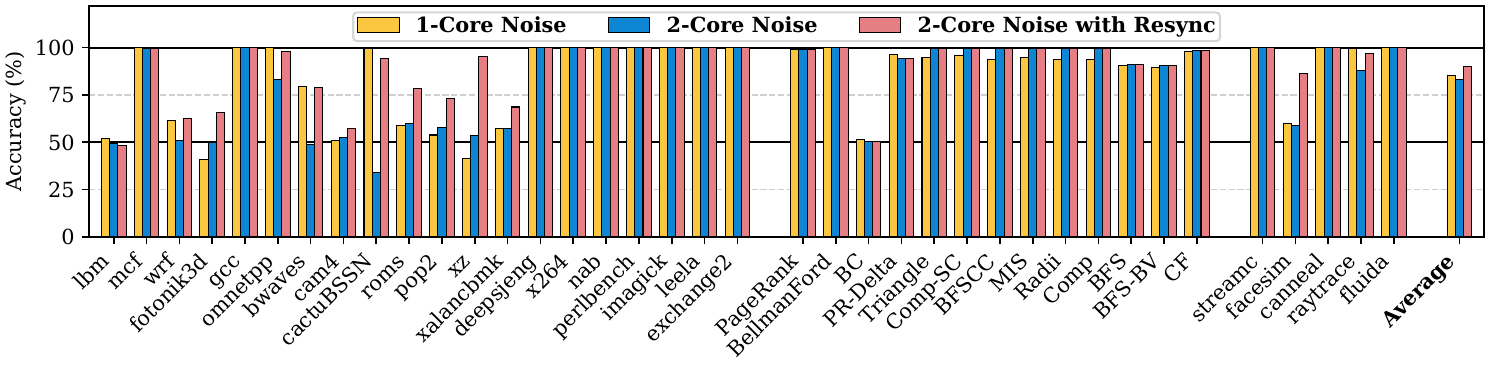}
    \vspace{-1em}
    \caption{Accuracy of the covert channel when running SPEC2017, LIGRA, and PARSEC workloads in parallel with the covert channel. The average accuracy of the channel with one~(1-Core Noise) and two~(2-Core Noise) extra workloads running in parallel to add noise is 85\% and 83\%, respectively. The accuracy with 2-core noise improves to 90\% if we perform a REF synchronization after every 100 bits are transmitted.}
    \Description{Accuracy of the Covert Channel with Noise}
    \label{fig:covch_noise}
\end{figure*}

\subsection{Results}
\begin{figure}[htb]
    \centering
    \includegraphics[width=\linewidth]{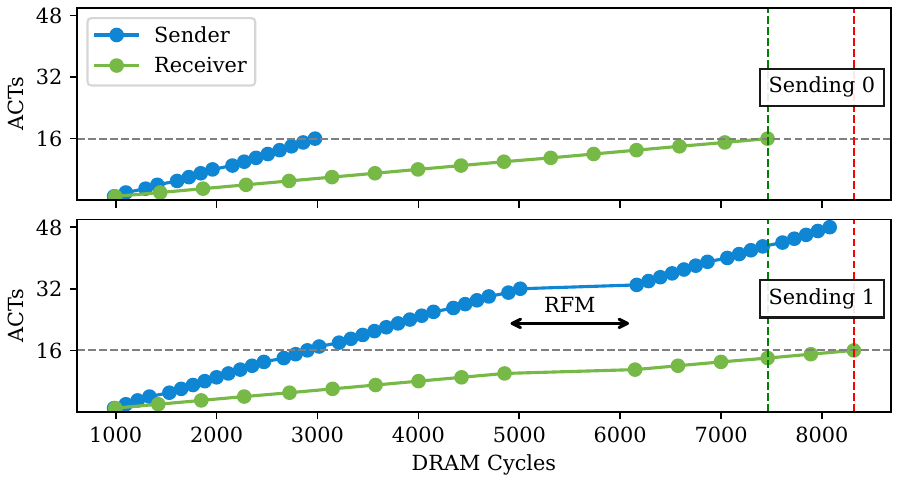}
    \vspace{-1em}
    \caption{The activations of the sender and receiver over time. While sending `1', the sender triggers \RFMabEND, which causes a significant delay in the response to the receiver.}
    \Description{Covert Channel Results}
    \label{fig:result}
\end{figure}

We run the sender and receiver gadgets on a 2-core simulated system with the setup described in \Cref{sec:methodology}. \Cref{fig:result} shows the time series of activations from both the sender and receiver when sending 0 and 1.

\parhead{Time Difference}
The time taken by the receiver to complete 16 loads when receiving `0' and `1' is 3109 ns and 3465 ns, respectively. The resulting time difference is 356 ns, which is very close to the expected value of 410 ns~(\tRFCEND). This time difference can be detected by using the \texttt{rdtsc} instruction.

\parhead{Speed and Bandwidth}
The steady state of our covert channel relies on the \REF command decrementing the \RAACtr of all the banks in the rank by \RAAIMTEND/2. Hence, the covert channel has a maximum bandwidth of 1 bit every \tREFI cycles. Given \tREFIEND=3900 ns, the covert channel has a speed of 31.3 KB/s. Finally, since every sub-channel in DDR5 can run independently, for our setup with 2 sub-channels, the covert channel has a total bandwidth of 62.6 KB/s.

\parhead{Accuracy}
For this experiment, the sender and receiver gadgets run in a noiseless environment. This results in a 100\% accuracy for the covert channel. Next, we evaluate the sensitivity of the covert channel to noise in the system.


\subsection{Sensitivity to Noise in the System}
\Cref{fig:covch_noise} shows the accuracy of the covert channel when running SPEC2017~\cite{SPEC2017}, LIGRA~\cite{shun:ligra} and PARSEC~\cite{bienia2008parsec} workloads in parallel with the covert channel. We evaluate the accuracy of the covert channel in two setups: 1-core noise and 2-core noise. The sender sends alternating bits~(0101) to the receiver, and the receiver runs for 100 million instructions and measures the accuracy of the covert channel.  We observe that the average accuracy of the covert channel is 85\% and 83\% for 1-core noise and 2-core noise setup, respectively. Finally, while most workloads do not affect the accuracy of the covert channel, some workloads reduce the accuracy of the covert channel to 50\%, which is the same as random guessing.

\parhead{Root Cause of Reduced Accuracy}
We find that the sender and receiver get de-synchronized during periods of high memory activity in the parallel running workloads. As a result, any bits inferred by the receiver after de-synchronization are incorrect, leading to reduced accuracy.

\parhead{Improving Accuracy in a Noisy Setup}
To improve the accuracy of the covert channel in the noisy setup, we simply re-synchronize the sender and receiver to \tREFI after every 100 bits. This ensures that the sender and receiver get re-aligned to \tREFI if they get desynchronized due to noise. \Cref{fig:covch_noise} shows that the accuracy of the covert channel improves to 90\% for 2-core noise setup after re-synchronization.

\subsection{Countermeasures}
\label{sec:mitig_covch}
Memory isolation techniques such as bank and bus partitioning, which are used to mitigate existing memory-based covert channels, do not mitigate our covert channel.   

\parhead{Sub-Channel Partitioning}
One way to prevent our covert channel is to perform memory isolation using sub-channel partitioning. Unlike bank partitioning, sub-channel partitioning assigns all the memory within a sub-channel to a single application. A sender and a receiver application running on such a system will be isolated from each other since they will be assigned to different sub-channels that operate independently.

\parhead{Limitations}
Sub-channel partitioning is not always feasible since a system only has a limited number of sub-channels. Most commercial server-grade systems only support up to 16 sub-channels. This will limit the maximum number of applications that can run concurrently. Additionally, this will underutilize the resources available in the memory system.
\section{Denial-of-Service using RFM}
\label{sec:dos}
In this section, we describe how an attacker can exploit the RFM interface to reduce the performance of the victim application and potentially launch a Denial-of-Service (DOS) attack. We exploit the side effect of the RFM interface that allows one bank to induce a slowdown in all the other banks in the rank. An \RFMab command stalls all the banks for roughly 12\% of the available \tREFI period. Thus, an adversarial workload can trigger multiple \RFMab within a tREFI and reduce the available throughput for all the co-resident victim applications. In this section, we first describe an attack pattern that can trigger multiple \RFMabEND, and next, we evaluate the slowdown caused by this attack pattern when run along with co-resident applications. Finally, we discuss and evaluate a possible countermeasure against the DOS attack pattern.

\begin{figure}[htb!]
    \centering
    \includegraphics[width=0.9\linewidth]{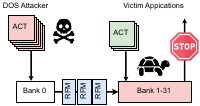}
    \caption{DOS Overview: The RFM interface allows one bank to reduce the performance of all the other banks.}
    \label{fig:dos}
    \Description{DOS Overview}
\end{figure}

\subsection{Attack Pattern}
The access pattern for the DOS attack is similar to that of the covert-channel. The attacker creates a self-evicting set and accesses it continuously in an infinite loop. Since the entire eviction set maps to different rows in the same bank, the \RAACtr of the attacker bank quickly reaches \RAAMMTEND. This triggers an \RFMab command, which stalls all the banks within the rank for about 410 ns~(\tRFCEND). After the \RFMab command is completed, the attacker keeps traversing the self-evicting set, resulting in even more \RFMabEND. This attack pattern can cause an average of 1.37 \RFMabEND/\tREFI for \RAAIMTEND=32 and 2.6 \RFMabEND/\tREFI for \RAAIMTEND=16. We devise an analytical model that can be used to predict the average \RFMabEND/\tREFI for a given \RAAIMTEND~(discussed in \nameref{appendix:a}).   


\begin{figure*}[htb!]
    \centering
    \includegraphics[width=\textwidth]{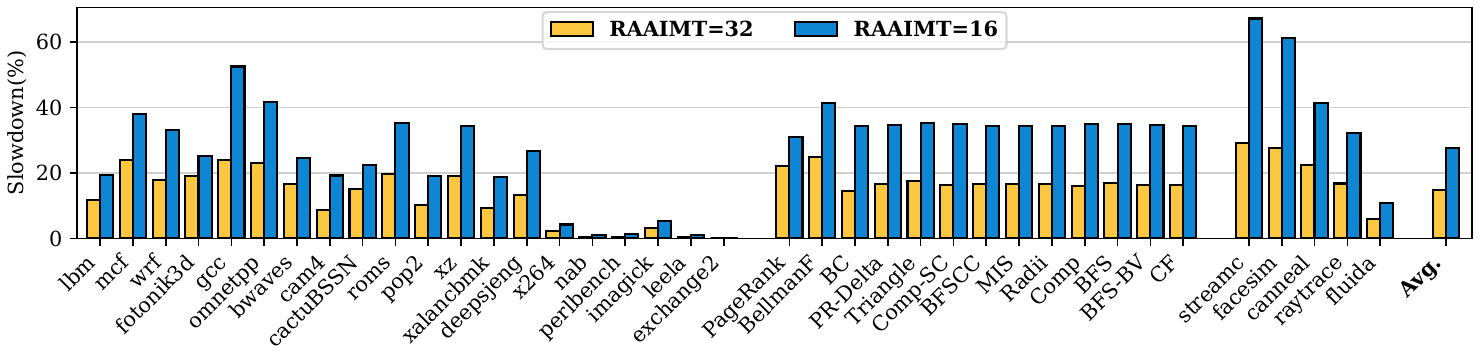}
    \caption{Normalized weighted slowdown caused by a DOS adversary triggering \RFMabEND. On average, the slowdown caused by the DOS adversary is 14.7\% for \RAAIMTEND=32 and 27.7\% for \RAAIMTEND=16. We see a worst-case slowdown of 29.3\% and 67\% for the \texttt{streamcluster} workload.}
    \Description{DOS Slowdown}
    \label{fig:dos}
\end{figure*}

\subsection{Evaluation}
\parhead{Setup}
In our 4-core system with 2 DRAM sub-channels, we run the attacker code on cores 0 and 1 and the victim workloads from \Cref{sec:methodology} on cores 2 and 3. The attacker code on core 0 continuously runs the DOS access pattern to a bank in sub-channel 0 and the attacker code on core 1 does the same to a bank in sub-channel 1. To quantify the slowdown, we measure the normalized weighted slowdown of the victim applications~\cite{eyerman2013restating}. 

\parhead{Results}
\Cref{fig:dos} shows the normalized weighted slowdown for all the workloads. The geometric mean of the slowdown across all the workloads is 14.7\% and 27.7\% for \RAAIMT of 32 and 16, respectively. We show that the DOS access pattern can result in maximum slowdowns of 29.3\% and 67\% for the \texttt{streamcluster} workload with \RAAIMT of 32 and 16, respectively. Thus, the side effects of RFM can allow an adversary to cause considerable slowdown in coresident applications.

\subsection{Countermeasures}
\label{sec:mitig_dos}
We design the following countermeasure to protect against the DOS attack pattern:

\parhead{Insight}
We observe that the number of \RFMab commands per \tREFI for benign applications is very low~(0.25 \RFMab per \tREFI for \RAAIMTEND=16). Therefore, limiting the number of \RFMab commands should not impact the performance of benign applications. However, this will prevent a DOS attacker from triggering continuous \RFMab commands and blocking the sub-channel for extended periods.

\parhead{Solution}
To limit the number of \RFMab commands per core, we limit the per-core activations to each bank~(per \tREFIEND). Thus, limiting the number of \RFMab commands triggered by each core. Our goal with this countermeasure is to limit the number of \RFMab commands to an average of 1 per \tREFI interval. As a result, we expect a minimal slowdown (less than 1\%) for benign applications and an upper limit on the slowdown when running the DOS attack~(less than 10\% per sub-channel). To reduce the impact of our countermeasure on benign applications with bursty access patterns, we allow unrestricted activations from each core if no RFM has been triggered in the last 16 \tREFI intervals. 


\parhead{Evaluation}
\Cref{tab:countermeasure} shows the relative slowdown after enabling the countermeasure. We observe that the impact of the countermeasure on benign applications is less than 1\%. However, on the other hand, the countermeasure reduces the slowdown caused by the DOS access pattern to almost half in the worst case. On average, the countermeasure limits the slowdown to less than 10\% per sub-channel. Thus, our proposed countermeasure can effectively limit the impact of the DOS attack.

\begin{table}[htb!]
    \centering
    \footnotesize
    \caption{Countermeasure Evaluation: Slowdown for benign and DOS applications (with and without countermeasure).}
\begin{tabular}{|c|c|c|c|c|c|}
    \hline
    \multirow{2}{*}{\textbf{RAAIMT}} &  \multirow{2}{*}{\textbf{Benign}} & \multicolumn{2}{c|}{\textbf{DOS (With Mitig.)}} & \multicolumn{2}{c|}{\textbf{DOS (No Mitig.)}} \\ \cline{3-6}
    &  & Max & Avg & Max & Avg \\ \hline \hline
    32 & 0.0\% & 16.45\% & 10.2\% & 29.3\% & 14.7\%\\ \hline
    16 & 0.48\% & 33\% & 14.2\% & 67\% & 27.7\%\\ \hline
\end{tabular}
    \label{tab:countermeasure}
\end{table}

\section{Discussion}
\label{sec:discussion}
\subsection{Impact of DRAM Address Mapping Function}
A different DRAM address mapping function should not impact the covert channel or the DOS attack. The attacker can reverse engineer the DRAM address mapping function~\cite{pessl2016drama}. Next, to evict the cache lines to the desired bank, the attacker can use the \texttt{clflush} instruction if the self-evicting set trick does not work.

\subsection{Covert-Channel using \RFMsbEND}
\parhead{\RFMab vs \RFMsbEND}
Unlike the \RFMab command which blocks all the banks within the rank, the \RFMsb command only stalls a bank set (the same bank across all bank groups). In our configuration, which consists of 4 banks per bank group and a total of 8 bank groups, the \RFMsb command will block 8 banks. A spy can exploit this partial stalling to create a covert channel similar to the one described in \Cref{sec:covch}.

\parhead{Speed of Covert Channel using \RFMsbEND}
The \RFMsb command is only allowed when Fine Grained Refresh (FGR) mode is enabled in DDR5~\cite{micron_ddr5}. In FGR mode, the \REF commands are issued every \tREFIEND/2 interval. This means that the covert channel can maintain the steady state and also transmit a bit every \tREFIEND/2 interval. The \RFMsb command requires 190 ns to execute. So, if the receiver is capable of detecting a time difference of 190 ns, the spy can potentially transmit two bits within a single \tREFI period using the \RFMsb command.

\section{Related Work}
\label{sec:related}
In this work, we focus on the security implications of the RFM interface. We have already discussed in-DRAM \rh defenses throughout the paper. In this section, we discuss other related work, especially those related to \rh and covert channels.

\subsection{Rowhammer}
\parhead{Tracking Aggressor Rows} Extensive research has been done to identify aggressor rows efficiently. This identification can either be done probabilistically or by counting activations to specific rows. PRA~\cite{kim2014architectural}, PARA~\cite{kim2014flipping}, MRLOC~\cite{MRLOC}, and ProHIT~\cite{PROHIT} are probabilistic approaches, while CRA~\cite{kim2014architectural}, CBT~\cite{cbT}, TWiCe~\cite{lee2019twice}, and Graphene~\cite{park2020graphene} count activations to specific rows. Prior works have also proposed exhaustive trackers that track activations to all rows in the DRAM array. These include CRA~\cite{kim2014architectural}, Hydra~\cite{qureshi2022hydra}, Panopticon~\cite{bennett2021panopticon} and PRHT~\cite{isscc23}. A recent revision to DDR5 specifications~\cite{JEDEC-PRAC} has extended the support for \emph{Per-Row Activation Counting} (PRAC). It also adds an \emph{ALERT-Back-Off} (ABO) interface, which allows the DRAM to alert and pause the memory controller when the ACTs to a row exceed an Alert threshold. Similar to RFM, ABO can also be used to build a covert channel and invoke a DOS access pattern.

\parhead{Mitigating Victim Rows} Throughout the paper, we assume that mitigation for \rh is performed by refreshing the neighboring victim rows. However, this can result in side effects such as Half-Double~\cite{HalfDouble}. Prior work has also looked at alternative mitigation techniques. For example, Blockhammer~\cite{yauglikcci2021blockhammer} limits the rate of activations to an aggressor row; this, in turn, restricts any row from reaching \TRH activations within the refresh period. Next, row-migration techniques, such as RRS~\cite{saileshwar2022RRS}, AQUA~\cite{saxena2022aqua}, SRS~\cite{SRS}, and SHADOW~\cite{wi2023shadow}, perform mitigation by moving an aggressor row to another location in memory. Thus limiting the maximum number of activations to the same physical row.

\parhead{Software-Based Defenses} Although software-based defenses~\cite{aweke2016anvil,van2018guardion,konoth2018zebram,bock2019rip} can prevent Rowhammer, they often require knowledge of DRAM properties that may be proprietary or not easily available to software. CATT~\cite{catt} tests DRAM cells and blacklists pages, which can cause significant loss of memory capacity at low \TRHEND. GuardION~\cite{van2018guardion} inserts a guard row between data of different security domains. ZebRAM~\cite{konoth2018zebram} and RIP-RH~\cite{bock2019rip} provide isolation by keeping the kernel space and user space(s) in different parts of DRAM.

\subsection{Covert-Channels}
Existing covert channels mainly exploit two shared hardware resources, caches and memory.

\parhead{Cache-based} Flush+Reload~\cite{yarom2014flush+} exploits the timing difference between a cache hit or miss in the LLC for access to a shared address. To send a bit, the sender either loads the shared address or performs a \texttt{clflush} on it. Flush+Flush~\cite{gruss2016flush+} is a variant that relies on the timing difference of the \texttt{clflush} instruction between cached and uncached lines. Prime+Probe \cite{osvik2006cache,liu2015last} is another cache-based covert channel that measures the timing difference between an LLC hit or miss. Unlike Flush+Reload and Flush+Flush, it does not require a shared address between the sender and receiver. Instead, it uses eviction sets to remove a cache line from the LLC.

\parhead{Memory-based} Xiao et al.~\cite{xiao2013security} exploits memory deduplication in hypervisors to create a covert channel between two VMs. Memory deduplication reduces the memory footprint by merging memory pages with identical content across VMs. This results in a timing difference: a write to a merged page is slower than a write to a non-merged page. Wu et al.~\cite{wu2014whispers} build a memory bus contention-based covert channel between two VMs by using the atomic instructions in x86 to lock the memory bus. Pessl et al.~\cite{pessl2016drama} developed a covert channel that relies on timing differences between a row buffer hit and a row buffer miss in DRAM.
\section{Conclusion}
\label{sec:conclusion}
Lowering \rh thresholds has increased the time needed to perform mitigative action. Existing mitigation approaches for in-DRAM defenses that steal time from REF do not scale at lower \TRHEND. To address this, JEDEC added the RFM interface to DDR5/LPDDR5 specifications. The RFM interface allows the memory controller to perform mitigative actions outside the \REF window. So far, researchers have used the RFM interface for its intended purpose of building defenses that can tolerate low thresholds. In this work, we analyze the potential security implications of the RFM interface. We demonstrate that the addition of the RFM interface to the DRAM protocol introduces new side effects that allow one bank to interfere with the operation of another, opening the door for its potential misuse. We use this side effect to build a new memory-based covert channel that has a bandwidth of 31.3 KB/s per sub-channel. Unlike existing memory-based covert channels, our channel cannot be mitigated using memory isolation techniques like bank and bus partitioning. Next, we also build a DOS attack pattern that uses the side effects of RFM to reduce the performance of coresident applications by up to 67\%.

\appendix
\section*{Appendix A}
\label{appendix:a}
We build the following analytical model to estimate the approximate number of \RFMab that can be launched within a \tREFI interval. 

\parhead{Analytical Model}
Let $\mathbf{nRFM}$ be the number of \RFMab triggered within s \tREFI window. Our DOS pattern continuously triggers activations to one of the banks, each of which consumes $\tRC$ time. Therefore, the total time \tREFI in a refresh cycle is equal to the sum of time taken by \REF command, the \RAAIMTEND/2 activations available after \REFEND, the time taken by \RFMab commands and the activation budget available after each \RFMab command~(see \Cref{eq:nRFM1}).
\begin{align}
    \tREFI &= \tRFC + \tRC \times \RAAIMTEND/2 + \mathbf{nRFM} \times \tRFCEND \nonumber \\ 
    &\qquad + \mathbf{nRFM} \times \RAAIMTEND \times \tRC
    \label{eq:nRFM1}
\end{align}
So, the number of \RFMab commands that can be triggered in a \tREFI window can be calculated as shown in \Cref{eq:nRFM2}.
\begin{align}
    \Rightarrow \mathbf{nRFM} &= \frac{\tREFI - \tRFC - \tRC \times \RAAIMTEND/2}{\RAAIMTEND \times \tRC + \tRFCEND}
    \label{eq:nRFM2}
\end{align}

\begin{figure}[ht]
    \centering
    \includegraphics[width=0.75\linewidth]{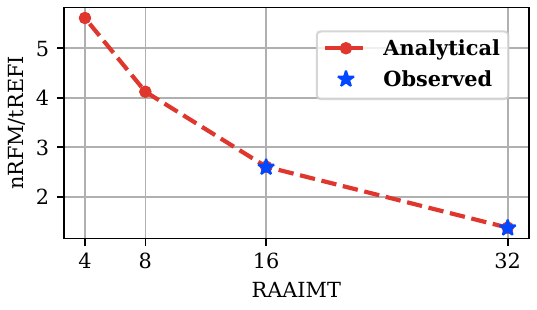}
    \caption{Average number of RFM triggered by the DOS pattern. The analytical and observed values closely align.} 
    \Description{DOS Design}
    \label{fig:dos_analytical}
\end{figure}

\noindent As shown in \Cref{fig:dos_analytical}, the analytical model closely aligns with the simulation results.

\parhead{Maximum Slowdown in Coresident Applications}
\RFMab roughly consumes 11.7\% of time within a \tREFI window and with \RAAIMTEND=32 and 16, the attacker can trigger a 1.37 and 2.6 \RFMab commands within a \tREFI window, respectively. Thus, we can expect a maximum slowdown of 16\% and 31\% with \RAAIMTEND=32 and 16, respectively.

\section*{Acknowledgements}
We thank the reviewers of ASPLOS-2025 for their valuable feedback and suggestions.


\bibliographystyle{plain}
\bibliography{refs}


\end{document}